# Effect of Foam Insertion in Aneurysm Sac on Flow Structures in Parent Lumen: Relating Vortex Structures with Disturbed Shear

Pawan Kumar Pandey[1], Malay Kumar Das[2]

Department of Mechanical Engineering, Indian Institute of Technology Kanpur

Kanpur 208016, Uttar Pradesh, India

## ABSTRACT

Numerous studies suggest that disturbed shear, causing endothelium dysfunction, can be related to neighboring vortex structures. With this motivation, this study presents a methodology to characterize the vortex structures. Precisely, we use mapping and characterization of vortex structures' changes to relate it with the hemodynamic indicators of disturbed shear. Topological features of vortex core lines (VCLs) are used to quantify the changes in vortex structures. We use the Sujudi-Haimes algorithm to extract the VCLs from the flow simulation results. The idea of relating vortex structures with disturbed shear is demonstrated for cerebral arteries with aneurysms virtually treated by inserting foam in the sac. To get physiologically realistic flow fields, we simulate blood flow in two patient-specific geometries before and after foam insertion, with realistic velocity waveform imposed at the inlet, using the Carreau-Yashuda model to mimic the shear-thinning behavior. With homogenous porous medium assumption, flow through the foam is modeled using the Forcheimmer-Brinkmann extended Darcy model. Results show that foam insertion increases the number of VCLs in the parent lumen. The average length of VCL increases by 168.9% and 55.6% in both geometries. For both geometries under consideration, results demonstrate that the region with increased disturbed shear lies in the same arterial segment exhibiting an increase in the number of oblique VCLs. Based on the findings, we conjecture that an increase in oblique VCLs is related to increased disturbed shear at the neighboring portion of the arterial wall.

KEYWORDS: Cerebral aneurysm; Hemodynamics; Vortex structures; Vortex core line; Disturbed shear; Shape memory polymer foam.

[1] Email- pawansut@iitk.ac.in, https://orcid.org/0000-0003-0145-3338
[2] Professor, Email- mkdas@iitk.ac.in, Ph- +91-512-259-7359, https://orcid.org/0000-0002-7971-4166

## ABBREVIATIONS

| | | | |
|---|---|---|---|
| WSS | Wall shear stress | VCL | Vortex core line |
| TAWSS | Time-averaged wall shear stress | ΔTAWSS | Change in time-averaged wall shear stress |
| TASWSS | Time-averaged secondary wall shear stress | ΔTASWSS | Change in time-averaged secondary wall shear stress |
| OSI | Oscillating shear index | ΔOSI | Change in oscillating shear index |
| SMP | Shape memory polymer | UPL | Upstream parent lumen |
| SBPL | Sac base parent lumen | DPL | Downstream parent lumen |
| EC | Endothelial cell | PLC | Parent lumen centerline |
| PC | Principal cord | | |

## 1. INTRODUCTION

Aneurysm often occurs at arterial sites having bifurcations, recirculation zones, and high oscillating shear zones [1]. These sites usually correspond to excessively high, low, or oscillating fluid loading acting on the endothelial cell (EC) layer. Existing literature suggests that these sites' neighboring hemodynamics act as a critical activator of the biochemical, cellular pathway responsible for wall dilation [1]. Prime hemodynamic indicator for aneurysm stability, i.e., wall shear stress, strongly correlates with the vortex dynamics at the aneurysm neck [2]. Although the debate is still not settled on how hemodynamics influences the tissue remodeling of arterial walls. Multiple mechanisms and indicators (wall shear stress and its derivatives) are reported as responsible for the aneurysm growth. Low and high wall shear stress and increased flow disturbance in the region are often associated with wall degeneration sites.

Reduction of intra-aneurysmal flow by separating the aneurysm sac from blood flow in the parent lumen is the primary goal of all aneurysm treatment procedures [3]. In surgical clipping, the aneurysm sac is disconnected from the parent artery by inserting a metal clip at the aneurysm neck. On the other hand, endovascular approaches such as coil filling attain it by reducing the flow inside the sac [4]. Low-flow hemodynamics inside the sac enables clot formation leading to subsequential vascular remodeling [5]. Despite widespread usage, endovascular coiling procedures have several disadvantages, such as coil compaction, dislocation, and inflammation [6]. Recent developments in polymer science pose shape memory polymer (SMP) foam as a viable alternative of coils for sac embolization. SMP foam is reported to show better performance than coils in terms of thrombus formation and biocompatibility [4,7,8]. SMP foams can be delivered in a temporarily compressed form inside the aneurysm sac and then triggered to unfold into its primary shape using external stimuli such as electrical resistance heating [9], ultrasound [10], lasers [11], and magnetic field [12]. Controllable porosity of foam allows for homogenous sac filling [9], which along with a suitably sculpted foam-lumen interface, is critically important in achieving improved hemodynamic loading at the neck [13,14]. Thus, due to significant potential advantages over coiling, foam-induced endovascular embolization is gaining traction as a new treatment technique for aneurysm treatment [9].

Even after filling the sac with porous media, aneurysm stability is not assured [6,15,16]. Several studies [17–21] find low wall shear stress responsible for aneurysm growth and rupture, while many other studies find high wall shear stress [22–25] as responsible. Another study [26] suggests that instead of wall shear stress, its derivatives serve a crucial role in disease initiation. Saqr et al. [27], in their review article, argued that the choice of wall shear stress and its derivatives as a hemodynamic indicator to bridge flow dynamics with mechano-biology of EC is based on modest reasoning and may prove insufficient in complex cases. One of their key argument that motivates this study is that endothelial dysfunction corresponds to WSS of disturbed flow rather than the uniform flow. They further argue that WSS is insufficient to justify the EC response of flow directionality [28,29] and flow structures. Meng et al. [30] reported a similar conflict of findings regarding high and low WSS. Arzani and Shadden in [31] suggested a revisit to the WSS characterization approach and its correlation with the diseases. They further add that WSS's vectorial nature and its gradients increase the data dimensionality and possible quantification of the hemodynamic indicators. EC(s) can sense their pulsating flow environment [32]. Le et al. [19] discussed the importance of vortex structure in the relation of flow pulsation with aneurysm stability and growth. Multiple studies report the presence of vortical structures close to the site of endothelial degeneration [17,22,33,34]. Vortex structures are often identified using iso-contours of some fundamental flow properties, and corresponding derived (velocity gradient) tensor fields and eigenvectors [35,36]. Although the velocity gradient is a local property, it can be used to identify global features such as vortex structures [37]. Various studies [19,38,39] have used Q-criterion and $\lambda_2$ criteria for vortex identification.

Identification and characterization of vortex structures is the critical requirement of its detailed analysis. While vortex structure identification using vorticity, Q-criterion, $\lambda_2$ and helicity provides a qualitative understanding using iso-surfaces; it cannot conveniently measure topological changes across different cases and waveform phases [2].

Patient-specific geometries also make it difficult to standardize the iso contour structures to compare across the different cases of vascular remodeling. Irregular iso-surface structures are also tricky for manual assessment. This work addresses the characterization issue of the vortex structures.

In the present work, we use vortex core lines (VCLs) to characterize the vortex structures. Using VCLs, Oeltze-Jafra et al. [40] presented a visual analysis of vortical structures in the cerebral aneurysm. Authors in [40,41] analyzed the embedded vortices phenomena occurring in cerebral aneurysms using a similar technique. VCL has some convenient topological information (more details in the 'Methods' section), which can be utilized to compare different cases and phases of pulsatile flow. Spiczak et al. [42] did a quantitative comparative study of blood flow in the aorta using features of VCL.

Our work presents a new approach to analyze porous media insertion effects on the vortex structures in the arterial lumen and its association with the disturbed shear. The present work studies the blood flow in a foam-inserted, patient-specific cerebral aneurysm subjected to the physiologically realistic blood-velocity waveform. The investigative focus is on understanding and characterization of vortex structures in the parent lumen, close to the aneurysm neck, before and after the foam insertion. The topological features of VCL are used for vortex structure characterization. Since most of the vascular remodeling happens over time significantly larger than a single pulse time period, this work characterizes vortex structures over the whole waveform period.

The present study is the first to address the comparative quantification issue of changes in vortex structures to the extent of our knowledge. The presented method enables us to relate the flow structures with disturbed shear, facilitating better assessment of flow structures' influence on endothelial dysfunction.

## 2. METHODS

A detailed account of various steps in numerical computations and analysis is provided in the following subsections. For brevity in presentation and discussion of results, some new terminology is also defined. The flow chart shown in Fig. 1 summarises all steps in numerical computations and data analysis.

### 2.1. Geometry

The present study uses two patient-specific geometries with a saccular aneurysm. These geometries are reconstructed from magnetic resonance angiography (MRA) images. The patient-specific geometry is obtained as an open-source mesh from the CISTIB laboratory at Universitat Pompeu Fabra of Barcelona [43]. Using open-source tools, i.e.,

GIMIAS and VMTK, geometries are constructed as shown in Fig. 2(a) and 2(b) [43]. Next, meshes for these geometries are generated using a commercial grid generator, i.e., ICEM-CFD. The inlet diameter for both geometries is 4.5 mm. Additionally, both geometries have the same upstream vasculature up to the aneurysm sac. There is a single inlet in both geometries; while geometry-1 has a single outlet, geometry-2 has two outlets. As part of the treatment procedure, SMP foam is inserted in the aneurysm sac, and its effect on three segments of the parent lumen is investigated. Based on the position 'd', measured along the centerline, from the center of the planar slice at the upstream lip (slice-2 in Fig. 2(c)), three segments of the lumen are defined as follows (for visual illustration, see Fig. 2(c)):

a) Upstream parent lumen (UPL): $-0.004 \leq d < 0.0$; segment of parent lumen sandwiched between slice-1 and slice-2, shown in Fig. 2(c).
b) Sac base parent lumen (SBPL): $0.0 \leq d < 0.004$; segment of parent lumen sandwiched between slice-2 and slice-3, shown in Fig. 2(c).
c) Downstream parent lumen (DPL): $0.004 \leq d < 0.008$; segment of parent lumen sandwiched between slice-3 and slice-4, shown in Fig. 2(c).

### 2.2. Governing Equations and Boundary Conditions

In this work, we model blood flow as incompressible, single-phase, and non-Newtonian. Equations (1), (2), and (3) are corresponding governing equations. SMP foam is modeled as a homogeneous, isotropic, and saturated porous medium subjected to incompressible flow. These assumptions allow us to model the flow through foam using the volume averaging approach. Pre-assessment of flow regime suggests that flow in foam ranges from creeping laminar to inertial laminar regime [4,44]. Therefore, to account for wall shear effects along with the inertial effect, flow in porous media is modeled using Darcy-Brinkmann-Forcheimmer Model [14,45]:

$$\nabla.\langle\vec{V}\rangle = 0 \tag{1}$$

$$\frac{\rho_f}{\varepsilon}\left[\frac{\partial\langle\vec{V}\rangle}{\partial t} + \langle(\vec{V}.\nabla)\vec{V}\rangle\right] = -\nabla\langle P\rangle^f + \frac{\mu}{\varepsilon}\nabla^2\langle\vec{V}\rangle - \frac{\mu}{K}\langle\vec{V}\rangle - \frac{\rho_f F\varepsilon}{K^{0.5}}\left[\langle\vec{V}\rangle.\langle\vec{V}\rangle\right]\vec{J} \tag{2}$$

In the rest of the lumen, as porosity $\varepsilon = 1$ and permeability $K \to \infty$, equations 1 and 2 reduce to the Navier stokes equation. In equation (2), F is the dimensionless inertia term coefficient and $\vec{J}$ is unit vector oriented along the velocity vector $\vec{V}$.

These equations are solved with no-slip boundary conditions for flow at the wall. A physiological velocity waveform (Fig. 2(d)) is imposed at the inlet to generate physiologically realistic flow pulsation inside the blood vessel. According to the inflow waveform's pulsating phase, the Womersley profile is applied at the inlet. For all cases (listed in Table-1), we have used a constant pressure boundary condition at the outlet. In present simulations, walls are treated as rigid and impermeable.

The shear-thinning behavior of blood becomes significant in arteries with a small diameter. Present problem setup requires a non-Newtonian viscosity model able to predict viscosity in clear and porous media both. Tosco et al. [46] suggested a modified approach of shear rate estimation which enables usage of generalized Newtonian models for porous media. Their approach can be extended to the Carreau-Yasuda model [47]. Capability to predict viscosity both in clear and porous media along with its property of stability and realistic predictions [48] makes the Carreau-Yasuda model (Eqn. (3)) our model of choice to mimic the blood rheology

$$\frac{\mu(\dot{\gamma}) - \mu_\infty}{\mu_0 - \mu_\infty} = [1 + (m\dot{\gamma})^n]^{-a}\ ; \quad (3)$$

$$\text{Inside clear lumen} \left\{ \dot{\gamma} = \left(\frac{1}{2} e_{ij} e_{ij}\right)^{\frac{1}{2}};\ e_{ij} = \frac{1}{2}\left(\frac{\partial u_i}{\partial x_j} + \frac{\partial u_j}{\partial x_i}\right) \right.$$

$$\text{Inside porous foam} \left\{ \dot{\gamma} = \alpha \frac{u_p}{\sqrt{K\varepsilon}} \right.$$

Where zero shear viscosity $\mu_0 = 1.6 \times 10^{-1}\ \mathrm{Pa \cdot s}$, infinite shear viscosity $\mu_\infty = 3.5 \times 10^{-3}\ \mathrm{Pa \cdot s}$, and other parameters $m = 8.2\ s, n = 0.64$, and $a = 1.23$. Inside clear lumen, the dynamic viscosity $(\mu)$ varies with the shear rate $(\dot{\gamma})$ estimate based on the second invariant of the strain-rate tensor $(e_{ij})$. However, inside the porous medium, the shear rate is estimated as the scaled ratio of volume-averaged velocity $(u_p)$ and foam's representative length scale $(\sqrt{K\varepsilon})$. Value of scaling factor $(\alpha)$, in expression for shear rate estimate, can be taken as unity for homogenous porous media [46]. Viscosity calculations in the clear lumen and inner region of foam are done as per equation (3). However, the approach suggested by Tosco et al. [46] does not account for wall bounding effects on the shear rate. Therefore, viscosity values are taken as infinite shear viscosity in near-wall, wall shear dominated region inside foam [14].

### 2.3. Numerical Solution

Governing equations are solved using an in-house solver written in C++. The solution procedure is based on the finite volume method (FVM). Geometries are discretized using an unstructured tetrahedra mesh. To get velocity and pressure by solving momentum and continuity equations, the solver uses a modified SIMPLE algorithm proposed by A.W. Date in [49]. The present method uses a colocated variable approach and pressure smoothing to eliminate pressure checkerboarding [50]. Convective terms are discretized using a hybrid upwind scheme combining first and second-order schemes. Diffusion terms are discretized using a second-order central-difference scheme. The discretized system of algebraic equations is solved using stabilized bi-conjugate gradient method (BiCGStab) with a diagonal pre-conditioner. Iterations at each time step are done until a relative residual of $10^{-4}$ in velocity and pressure components is achieved. To accurately capture the physiological pulsatile flow, simulations are carried out until cyclic repeatability is achieved for at least three consecutive waveform cycles. Results of last waveform cycle is analyzed and reported at the time interval of $10^{-2}$ seconds.

### 2.4. Flow Parameters

The density of blood is taken as 1050.0 kg/m$^3$. Inlet flow waveform (Fig. 2(d)) used in simulations has a frequency of 1.17 Hz yielding Womersley number Wo = 6.7. For porous media, porosity (ε) is taken as 0.735, permeability (K) is $1.539 \times 10^{-8}$ m$^2$ and inertia coefficient (F) is taken as 0.2 [51].

### 2.5. Vortex Structure Identification and Characterization

To quantify the topological changes in vortex structures due to foam insertion, we extract vortex core lines (VCLs) from the velocity field obtained from the simulations. Ease in the calculation of topological characteristics of VCL is its distinct advantage over other methods of vortex structures visualization. VCL extraction is based on an algorithm proposed by Sujudi and Haimes [52]. For a more detailed discussion on the algorithm, the reader can refer to the other works [52,53]. Nonetheless, we list all major steps of computations as follows:

a) Find the rate of deformation tensor D for each cell.

b) Proceed if D has one real eigenvalue ($\lambda_R$) and two complex eigenvalues ($\lambda_C$).

c) Find the planar velocity normal to an eigenvector of real eigenvalue.

$$\vec{w} = \vec{u} - (\vec{u}.\vec{n})\vec{n} \quad (4)$$

$\vec{n}$ is a normalized eigenvector corresponding to a real eigenvalue

d) For all cell faces, identify whether reduced velocity ($\vec{w}$ from the previous step) is zero.

e) Cells with at least two faces, having zero reduced velocity, have a vortex core line passing through them.

f) Connect neighboring cells identified from step (e) and get vortex core line(s).

The extracted VCL suffers from several issues, such as breakage and spurious VCL predictions. Grid size and numerical error are the primary cause of VCL breakage. A cutoff based on the representative length scale of the grid is used to fix the breakage in VCL. Additionally, we applied helicity-based filtering to eliminate spurious VCL.

In the present study, we assume that the vortex in the region of interest, i.e., parent lumen (between slice-1 and slice-4 in Fig. 2(c)), moves much slower than a typical representative fluid particle [53]. This assumption enables us to employ this method for each phase of the inflow waveform. This assumption is further found reasonable by the visualization of VCL over the complete waveform.

To verify the correctness of the VCL extraction procedure, an artificial and simple flow field is generated by imposing helical flow at the inlet of a straight tube. For this flow field, VCL coincides with the centerline of the straight tube. This provides a verification of the present methodology. The straight tube geometry along with helical streamlines and extracted VCL is shown in Fig. 3(a).

For quantification purposes, VCLs are characterized based on their geometrical properties [54]. Some of these quantities, in the present work's context, is illustrated in Fig. 3(b) and defined as follows:

i. **Principal Cord (PC):** Straight line connecting the endpoints of a VCL.

ii. **Cord-length (L):** It is defined as the total length of the principal cord. Cord-length is considered as the representative length of a VCL. In further discussions, we use terminologies 'VCL length' and 'Cord-length' interchangeably.

iii. **Angle (θ):** It is defined as the angle between the principal cord of a VCL and parent lumen centerline (PLC). It is considered as a measure of the orientation of VCL with respect to PLC.

iv. **Position (d):** It is defined as the distance of the principal cord's center, measured along PLC, from the lumen center at the upstream lip.

### 2.6. Hemodynamic Indicators

Many quantities based on the flow fields or vessel anatomy are reported to show a strong correlation with the site of arterial wall degeneration [3,55]. However, to relate with the pathophysiological influence of flow structures and to identify the disturbed shear sites, we use the following three hemodynamic indicators in the present work:

#### 2.6.1. Change in Time Averaged Wall Shear Stress (ΔTAWSS)

Wall shear stress is estimated as a tangential component of the traction vector (or its magnitude) on the vessel's endothelial cell layer. Traction vector ($\vec{T}$) relates the stress tensor ($\bar{\bar{\tau}}$) with the outward unit normal ($\hat{n}$) of the wall surface:

$$\vec{T} = \hat{n}.\bar{\bar{\tau}} \tag{5}$$

After removing the normal component, we get the tangential component of the traction vector:

$$\vec{S} = \vec{T} - \vec{N} \tag{6}$$

Where normal component: $\vec{N} = \vec{T}.\hat{n}$

The magnitude of the tangential component of the traction vector $|\vec{S}|$ is considered as the value of wall shear stress (WSS). WSS's average value calculated over the whole waveform is called time-averaged wall shear stress (TAWSS). In the present work, we report the change in time-averaged wall shear stress, i.e., ΔTAWSS, calculated as follows:

$$\Delta \text{TAWSS} = \text{TAWSS after foam insertion} - \text{TAWSS before foam insertion} \quad (7)$$

### 2.6.2. Change in Oscillating Shear Index (ΔOSI)

As a measure of directional change of wall shear stress, He and Ku [56] defined the Oscillatory shear index (OSI) for general three-dimensional geometry as follows:

$$\text{OSI} = \frac{1}{2}\left(1 - \frac{\left|\int_0^T \vec{S}\ dt\right|}{\int_0^T \left|\vec{S}\right|\ dt}\right) \quad (8)$$

In this work, we present the contours of change in OSI, which is defined as the following:

$$\Delta \text{OSI} = \text{OSI after foam insertion} - \text{OSI before foam insertion} \quad (9)$$

The values of OSI ranges from 0 (uni-directional shear) to 0.5 (utterly oscillating shear). Therefore, ΔOSI, representing the difference in OSI values [57], theoretically, can vary from -0.5 to +0.5. While negative values of ΔOSI indicate a reduction in shear stresses' oscillatory nature, positive values indicate the increase in the oscillatory nature of shear stresses.

### 2.6.3. Change in Time Averaged Secondary Wall Shear Stress (ΔTASWSS)

To estimate shear stress exclusively due to azimuthal circulation of lumen flow, we define secondary wall shear stress (SWSS). The method of calculating SWSS is similar to WSS with the only difference of modified velocity field, which is calculated by deducting axial velocity from the actual velocity vector. At any point, axial velocity is calculated as the dot product of the velocity vector and unit normal at the closest point on the parent lumen centerline (PLC). In the subsequent sections, we report the change in time-averaged secondary wall shear stress (ΔTASWSS), which is defined as follows:

$$\Delta \text{TASWSS} = \text{TASWSS after foam insertion} - \text{TASWSS before foam insertion} \quad (10)$$

Here, TASWSS is the temporal average of SWSS taken over the entire inflow waveform.

## 2.7. Validation and Grid Independence

The solver has been extensively validated for steady and unsteady flows through various geometries [58], including pulsatile flows through clear and porous media [14]. The present work includes validation of pulsatile flow through a curved tube with a 90-degree bend [59]. The validated velocity profiles are shown in Fig. 3(c), and 3(d), and the corresponding curved tube is shown in Fig. 3(e). For present geometry and mesh, a grid independence study is already shown in our previous published work [60].

## 3. RESULTS

Time-lapse plots and time-averaged plots are the two approaches used in this work to present results for the whole waveform (shown in Fig. 2(d)). Time-lapse plots simultaneously present VCL data over complete waveform at a time interval of 0.01 seconds. Since the waveform's time period is 0.86 seconds, time-lapse plots include data points for 87-time instants. Time-averaged plots exhibit the data averaged over the time period of the inflow waveform. VCL and its topological characteristics are presented using time-lapse plots. On the other hand, the time-averaged plot is used to present contours of hemodynamic indicators defined in the previous section.

Figures 4(a) and 4(b) show the time-lapse plots of VCLs before and after foam insertion. Fig. 4(a) and 4(b) show that foam insertion causes VCLs to vanish from inside the sac, although the number of VCLs increases in the parent lumen. Figures 4(c) and 4(d) show the Q-criterion iso-contours along with VCLs, at waveform's systolic peak, before and after foam insertion, respectively.

Figure 5 shows the time-lapse scatterplots of VCLs' position and angle (for definitions, refer to 'Methods'). Figures 5(a) and 5(c) show position-angle (PA) scatterplot for before and after foam insertion in geometry-1, respectively. Similarly, Figures 5(b) and 5(d) show PA scatterplots for before and after foam insertion in geometry-2, respectively. Close to the upstream lip ($d = 0.0$), data points are nearly uniformly distributed over the complete range of angles. So, we infer that in the vicinity of the upstream lip, VCLs have no preferred orientation and appear in nearly all orientations. This behavior is visible in all four cases. In contrast, within DPL ($0.004 \leq d < 0.008$), VCLs make a small angle with PLC; however, this angle increases after foam insertion. In SBPL ($0.0 \leq d < 0.004$), compared to before foam insertion, after foam insertion, there is a significant increase in the number of VCLs appearing over the waveform's time period.

Position and cord-length (PL) scatterplots are shown in Fig. 6. In both geometries, foam inserted cases (Fig. 6(c) and 6(d)) show an increase in VCLs' cord-length near downstream lip ($d = 0.004$). This increase in VCL cord-length due to foam insertion is higher in geometry-1 than geometry-2. While before foam insertion, maximum VCL cord-length is higher in geometry-2 than geometry-1; after foam insertion, geometry-1 has longer VCLs than geometry-2. Scatterplots for geometry-2 show more numerous and smaller VCLs in DPL after foam insertion than before foam insertion case. In contrast, scatterplots of geometry-1 show a significant increase in VCL cord-length, in DPL, due to foam insertion (Fig. 6(a) and 6(c)).

Angle and cord-length (AL) scatterplots are shown in Fig. 7. Data points in these scatterplots, for both geometries without foam insert (Fig. 7(a) and 7(b)), exhibit no correlation between angle and cord-length of VCL. However, after foam insertion, longer than average VCLs are more likely to make small oblique angle (~0.5 radians and ~2.5 radians) with PLC (Fig. 7(c) and 7(d)).

Table 2 lists the average cord-length of VCLs, in parent lumen, for both geometries before and after the foam insertion. Before foam insertion, the average VCL cord-length of both geometries is in close range. After foam insertion, the average length shoots up for both geometries. A rise of 168.9% and 55.6% in average VCL cord-length is observed for geometry-1 and geometry-2, respectively.

Figure 8 presents histograms of percentage changes in the number of VCL against cord-length and angle due to foam insertion in both geometries. These plots are shown separately for UPL, SBPL, and DPL regions. For brevity, VCL with a cord-length smaller than 0.0045m is referred to as of small size; VCL with cord-length larger than 0.0045m and smaller than 0.0105m is considered to be of moderate size, and VCL with length larger than 0.0105m is treated as large VCL. The orientation of VCL is classified as aligned [$(0 \leq \theta < 0.315)$ & $(2.835 \leq \theta \leq \pi)$], oblique [$(0.315 \leq \theta < 1.26)$ & $(1.89 \leq \theta < 2.835)$] and normal [$1.26 \leq \theta < 1.89$] with respect to the PLC. Histograms in Fig. 8 map the changes in VCLs using ten bins, while table 3 lists the same data as per defined length (small, moderate, and large) and orientation (aligned, oblique, and normal) types.

In *UPL* (Fig. 8(a) & Table 3), aligned VCLs increase by 14.9% and decrease by 6.7% in geometry-1 and geometry-2, respectively. Moreover, oblique VCLs increase by 20.7% and 23.1% in geometry-1 and geometry-2, respectively. On the other hand, normal VCLs decrease by 11.1% and 8.9% in geometry-1 and geometry-2, respectively. Regarding the change in cord-length of VCLs (Fig. 8(b) & Table 3), small VCLs are observed to increase by 24.5% and 7.4% in geometry-1 and geometry-2, respectively. However, no changes were observed for moderate and large-sized VCLs.

In *SBPL* (Fig. 8(c) & Table 3), aligned VCLs increase by 20.7% and 12.2% in geometry-1 and geometry-2, respectively. Additionally, oblique VCLs increase by 61.7% and 22.1% in geometry-1 and geometry-2, respectively. Furthermore, normal VCLs also increase by 2.7% and 5.1% in geometry-1 and geometry-2, respectively. Regarding the change in cord-length of VCLs (Fig. 8(d) & Table-3), small VCLs see an increase of 67.4% and 31.1% in geometry-1 and geometry-2, respectively. Moderate-sized VCLs register an increase of 14.6% and 8.3% in geometry-1 and geometry-2, respectively. In contrast with small and moderate-sized VCLs, large-sized VCLs increase by a mere 3.1% in geometry-1, and no change is observed in geometry-2.

In *DPL* (Fig. 8(e) & Table 3), aligned VCLs decrease by 2.7% and increase by 14.7% in geometry-1 and geometry-2, respectively. While normal VCLs do not record any significant changes, oblique VCLs increase by 11.1% and 9.6% in geometry-1 and geometry-2, respectively. Regarding the change in VCL numbers with respect to cord-length (Fig. 8(f) & Table 3), small VCLs decrease by 0.4% in geometry-1 and increase by 24.7% in geometry-2. In geometry-1, moderate and large-sized VCLs increase by 2.7% and 6.1%, respectively. On the other hand, no changes were observed for moderate and large-sized VCLs in geometry-2.

Figures 9(a) and 9(b) show contours of change in TAWSS due to foam insertion in the sac of both geometries. Both geometries show a negative ΔTAWSS region in SBPL sandwiched between neighboring regions of positive ΔTAWSS. Figures 9(c) and 9(d) show contours of ΔOSI due to foam insertion in both geometries. Significant positive ΔOSI values are observed in the SBPL. Figures 9(e) and 9(f) show the contour of ΔTASWSS for both geometries. Except for the aneurysm sac, only SBPL contains the negative ΔTASWSS region in both geometries.

## 4. DISCUSSION

Presented results show that foam insertion in the aneurysm sac significantly alters the hemodynamics in the parent lumen. The presence of porous media (SMP foam) inside the aneurysm sac forces streamlines to adjust, from upstream lumen shape to downstream lumen shape, in SBPL only. Foam insert effectively blocks out the sac for re-orientation of flow. So, flow is forced to adjust itself only within SBPL; this is expected to reflect on the flow structures and wall loading in the SBPL. Investigation on these hemodynamic changes was done in two ways: first, by measuring the topological changes in vortex structures and second by calculating ΔTAWSS, ΔTASWSS, and ΔOSI.

A large chunk of existing literature uses one or another vortex characterizing quantity (Q-criterion, λ-criterion, and Δ-criterion) based iso-surfaces for studying the vortex structures [2,19,34]. These methods provide a visual understanding of the flow field. However, while comparing different cases, the iso-surface-based method cannot conveniently quantify the topological differences in the vortex structures. A simple skeleton of vortex structure found through VCLs can easily quantify the differences between vortex structures of two different cases. In this work, we used this logic to measure the differences in vortex structures formed in parent lumen before and after foam insertion. As shown in Fig. 4(c) and 4(d), filaments of Q-criterion (obtained after several manual iterations) and VCLs portray a visually similar picture of vortex structures. VCL properties such as length and orientation can be used for quantitative topological characterization [42]. However, initially extracted VCL using Sujudi and Haimes' algorithm [52] suffer from several issues such as breakage and spurious predictions. Thus, after extraction, VCLs are processed for enhancement. Authors in [40] have discussed the various aspects of VCL enhancement. An additional filter based on the average helicity of VCL is used to filter the spurious VCL [61]. Scatterplots based on the angle, cord-length, and position (Fig. 5, 6, and 7) show that before foam insertion highest density of VCLs exists at the upstream lip. After foam insertion, the highest density of VCLs exists in the SBPL. Except for VCLs concentrated around the upstream lip, most of the VCLs are aligned with PLC before foam insertion. However, after foam insertion, a significant number of VCLs become oblique with the PLC. Normal VCLs are found to increase only in the SBPL region due to foam insertion in the sac. The largest combined increase of oblique and normal VCLs is seen in SBPL of both geometries. Change histograms proved to be a very useful tool to understand the influence of foam insertion on VCLs in UPL, SBPL, and DPL.

Various empirical evidence suggests that vortex structures exposed to the endothelium cell layer are responsible for endothelial dysfunction by altering the shear stresses at the wall [62–64]. Thus, to further advance our understanding of the mechano-biological influence of flow structures on the EC layer, we visually explored the spatial similarity between the topological changes in vortex structures and changes in hemodynamic indicators, i.e., ΔTAWSS, ΔTASWSS, and ΔOSI. Low TAWSS and high OSI are well-accepted hemodynamic indicators for disturbed shear [65]. *Regions of maximum decrease in TAWSS and an increase in OSI are localized in the SBPL (encircled in Fig.9), thus implying development towards disturbed shear in the SBPL.* We conjecture that the generation of disturbed shear in SBPL is due to an increase in the number of non-aligned VCLs in SBPL after foam insertion in the sac. Next, the TASWSS estimate provides information about shear stress experienced by the arterial wall due to azimuthal circulations. Contours of change in TASWSS, i.e., ΔTASWSS, is non-zero only in SBPL (except aneurysm sac).

The coincidence of regions of non-zero ΔTASWSS (Fig. 9(e) and 9(f)), negative ΔTAWSS (Fig. 9(a) and 9(b)), and positive ΔOSI (Fig. 9(c) and 9(d)), along with the increased number of oblique and normal VCL in SBPL (Fig. 8(c) and 8(d)) supports our conjecture. Figure 10 visually depicts the presented conjecture. Figure 10(a) shows a tube with one aligned VCL corresponding to a strong dominant helical flow. As suggested in [66], exposure of this kind of VCL to the surrounding wall can be considered physiologically positive. Figure 10(b) depicts a case where many VCLs are obliquely oriented. As found in the present work increased number of oblique VCL are related to disturbed shear on the wall. Thus, the VCL topology shown in Fig. 10(b) would lead to disturbed shear and subsequent endothelial degeneration and therefore represent adverse exposure.

A significantly higher increase in VCLs' cord-length and the number of oblique VCLs in geometry-1 than geometry-2 illustrates the effect of downstream vasculature (Table-2). However, further study with many more geometries would be required to understand the reasons for such behavior.

## 5. CONCLUSIONS

In this work, we investigated the effect of SMP foam insertion, in an aneurysm sac, on the flow structures in the parent lumen. Comparing vortex structures of different cases or different time instants is difficult using iso-surface plots of Q-criterion. We introduced a VCL characterization method, which enables us to compare the topology of vortex structures before and after foam insertion. Our analysis shows that, due to foam insertion in the sac, the vortex core lines (VCLs) in sac base parent lumen (SBPL) increase in number and become more oblique to the parent lumen centerline (PLC). Key hemodynamic indicators like TAWSS and OSI also predict the same region (SBPL) as the site developing disturbed shear. With our limited simulation results, we conjecture that an increase in the number of oblique VCLs, in a blood vessel leads to disturbed wall shear in the region. However, several more investigative studies are required to establish the proposed conjecture conclusively.

## ACKNOWLEDGEMENT

The High-Performance Computing facility at the Indian Institute of Technology Kanpur, India, is gratefully acknowledged.

## DECLARATIONS

**Conflicts of interest/Competing interests:** None

Table 1: Summary of cases studied in present work

| Cases | Aneurysm Sac | Geometry |
|---|---|---|
| **Case 1** | without foam insert | Geometry 1 |
| **Case 2** | with foam insert | Geometry 1 |
| **Case 3** | without foam insert | Geometry 2 |
| **Case 4** | with foam insert | Geometry 2 |

Table 2: Average length of VCL under aneurysm sac

| Average VCL length | Before foam insertion (m) | After foam insertion (m) | % Change |
|---|---|---|---|
| Geometry-1 | 0.0058 | 0.0156 | + 168.9% |
| Geometry-2 | 0.0063 | 0.0098 | + 55.6% |

Table 3: Summary of changes in the number of vortex core lines (VCL) over the time period of inflow waveform. VCLs are classified based on angle and cord-length; changes are reported for parts of parent lumen, i.e., upstream parent lumen (UPL), sac base parent lumen (SBPL), and downstream parent lumen (DPL)

| % Change in VCL | | Angle (θ) | | | Cord-length (L) | | |
|---|---|---|---|---|---|---|---|
| | | Aligned [(0≤θ<0.315) & (2.835≤θ≤π)] | Oblique [(0.315≤θ<1.26) & (1.89≤θ<2.835)] | Normal [1.26≤θ<1.89] | Small [L< 0.0045m] | Moderate [0.0045m ≤ L < 0.0105m] | Large [L ≥ 0.0105m] |
| Geometry-1 | UPL | +14.9% | +20.7% | -11.1% | +24.5% | 0.0% | 0.0% |
| | SBPL | +20.7% | +61.7% | +2.7% | +67.4% | +14.6% | +3.1% |
| | DPL | -2.7% | +11.1% | 0.0% | -0.4% | +2.7% | +6.1% |
| Geometry-2 | UPL | -6.7% | +23.1% | -8.9% | +7.4% | 0.0% | 0.0% |
| | SBPL | +12.2% | +22.1% | +5.1% | +31.1% | +8.3% | 0.0% |
| | DPL | +14.7% | +9.6% | +0.3% | +24.7% | 0.0% | 0.0% |

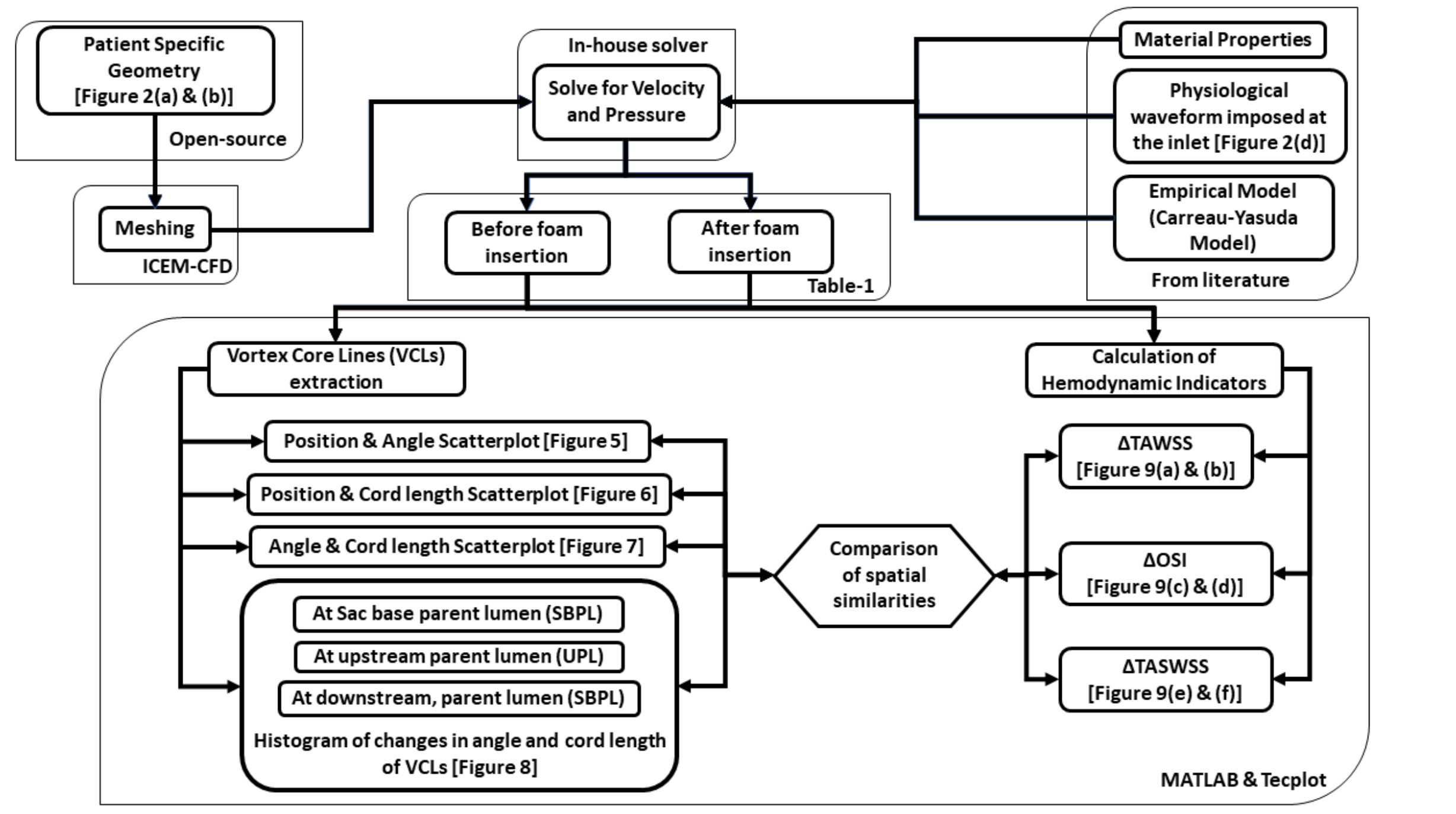

Figure 1: Schematic of computations and analysis undertaken in this work

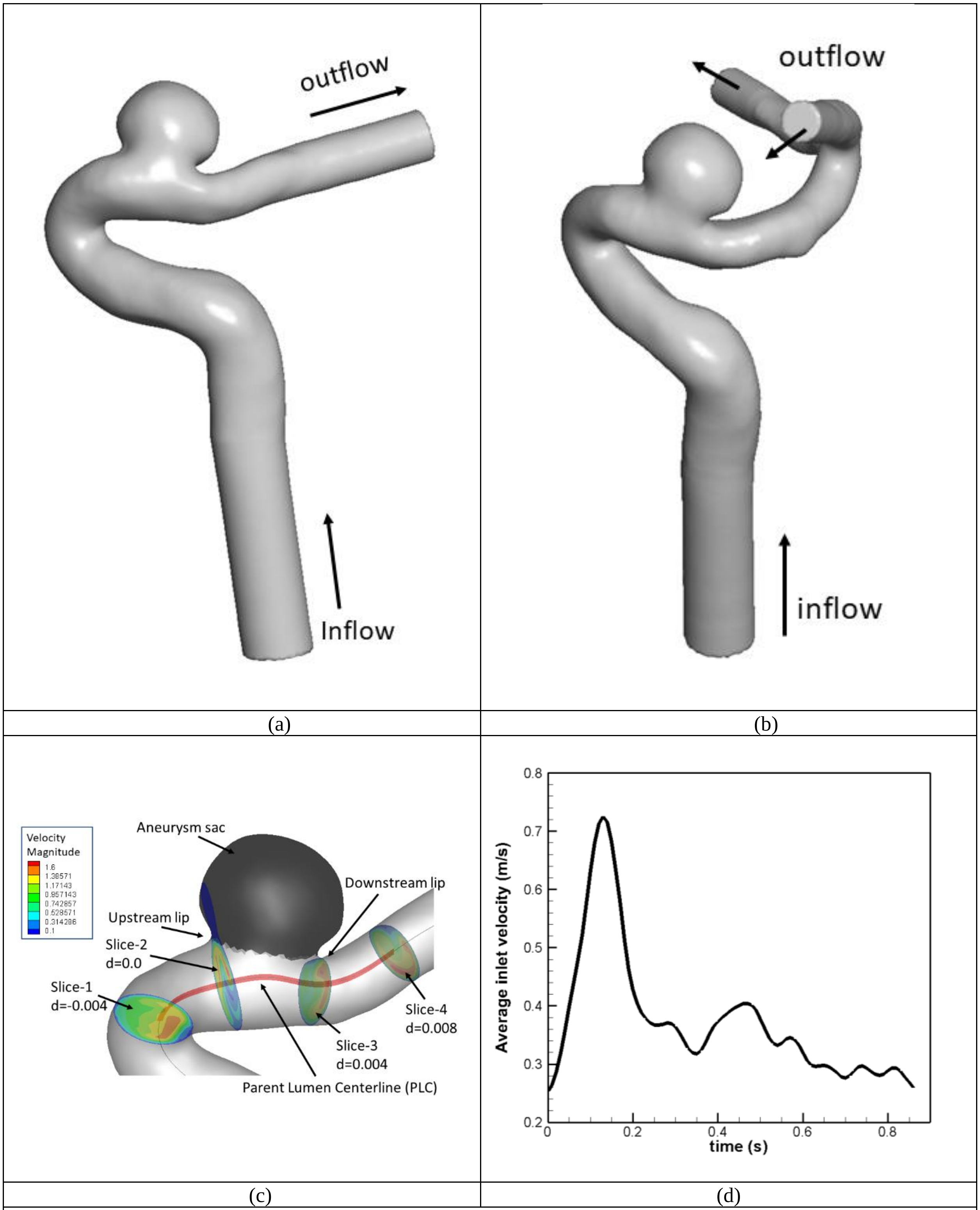

Figure 2: Patient-specific geometries: (a) geometry-1; (b) geometry-2; (c) Nomenclature and position 'd' details of region of interest in parent lumen (between slice-1 and slice-4) in which change in vortex structures due to foam insertion in the aneurysm sac (shown in the dark shade) is studied; slice-2 and slice-3 are taken at the upstream lip, and downstream lip of aneurysm, respectively; (d) Velocity waveform imposed at the inlet

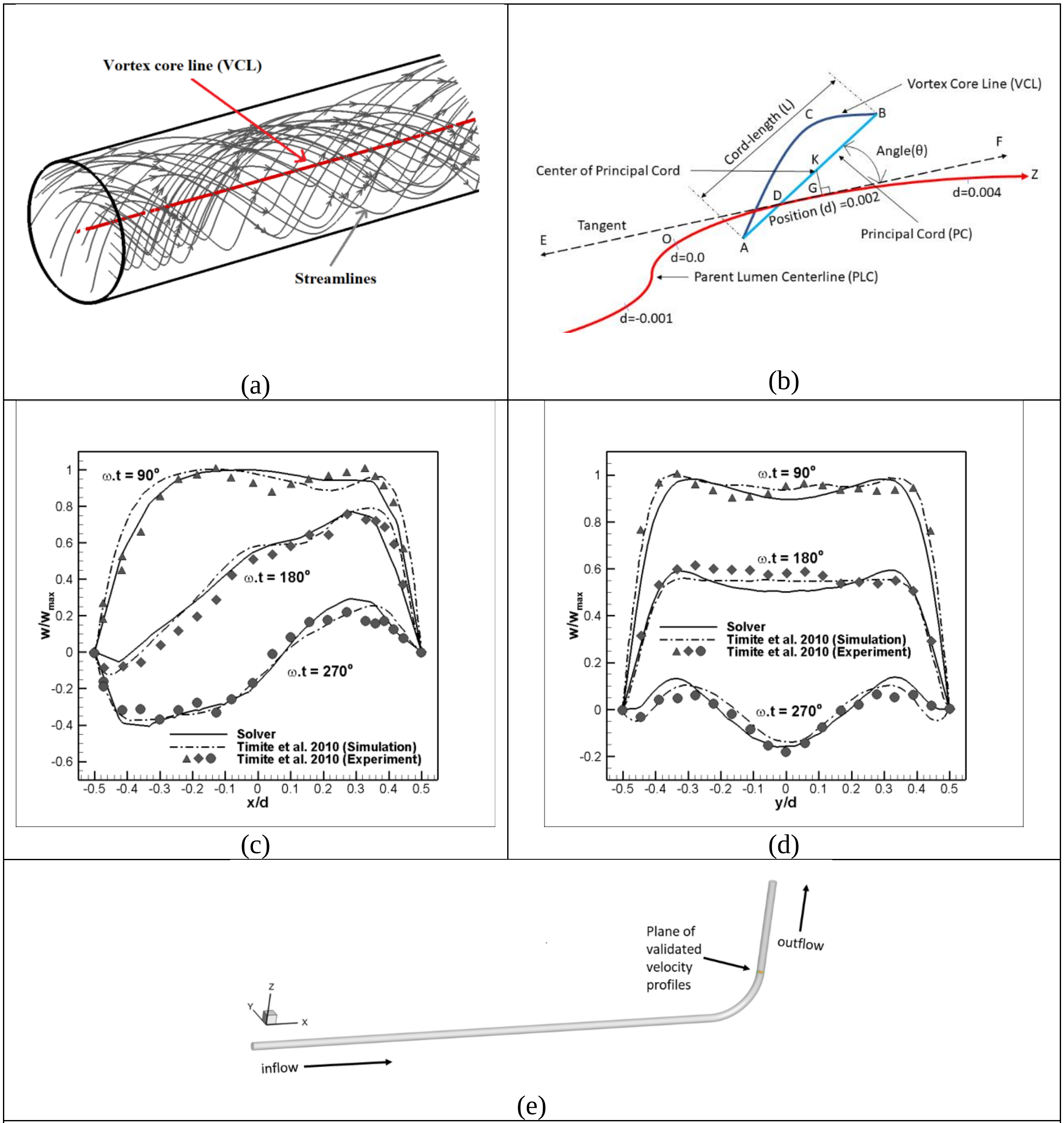

Figure 3: (a) Extracted vortex core line (VCL) for helical flow inside a straight tube coincides with the tube's centerline; (b) Schematic illustration of the calculation of VCL's geometrical properties such as cord-length, angle, and position with respect to parent lumen centerline (PLC); (c)-(d) velocity profiles, of three different phases, at the end of the 90-degree bend in the (e) curved tube validated with experimental and numerical results of Timite et al. [59]. In plots (c)-(d), $w/w_{max}$ represents the axial velocity non-dimensionalized with local maxima, $\omega$ represents the angular frequency, t represents the time instant, $\omega.t$ represents the phase angle corresponding to sinusoidal wavform imposed at the inlet of the curved tube

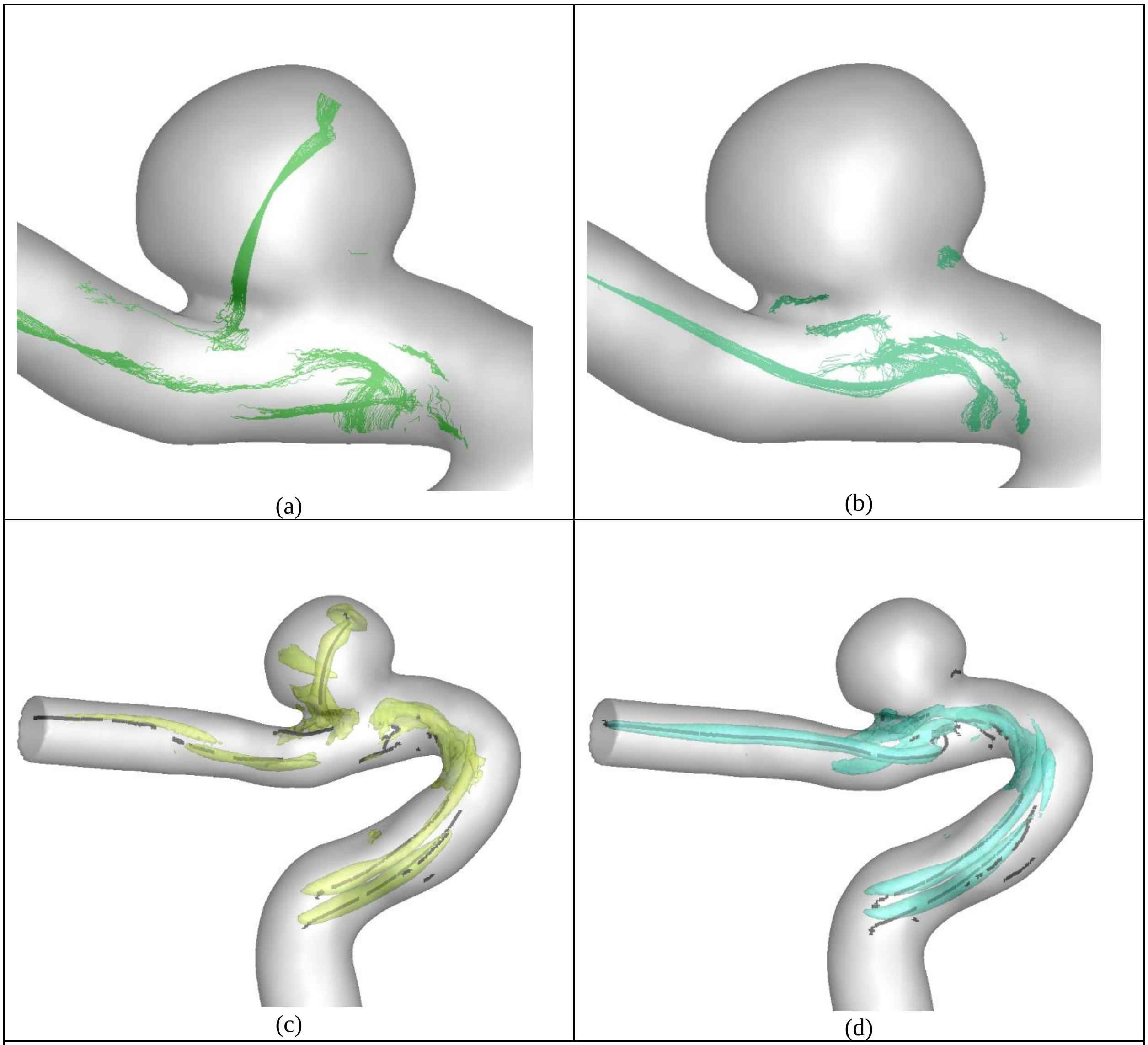

Figure 4: Time-lapse plot of vortex core lines (VCLs) in geometry-1 (a) before foam insertion (b) after foam insertion. The simultaneous plot of VCLs and Q-criterion iso-contours at systole peak (t = 0.13 s) for geometry-1 (c) before foam insertion (d) after foam insertion

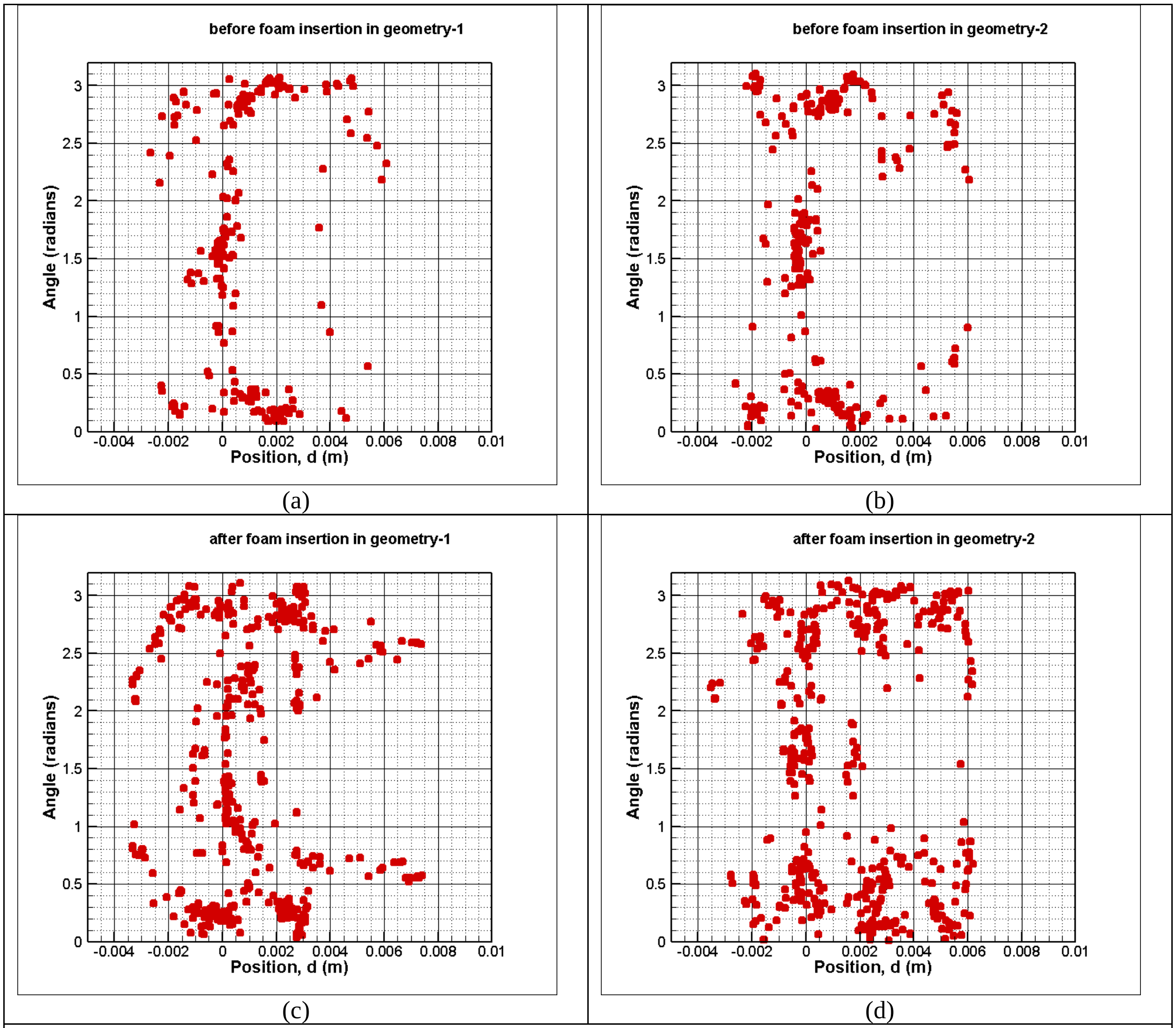

Figure 5: Vortex core lines' position and angle scatterplot for (a) before foam insertion in geometry-1 (b) before foam insertion in geometry-2 (c) after foam insertion in geometry-1 (d) after foam insertion in geometry-2

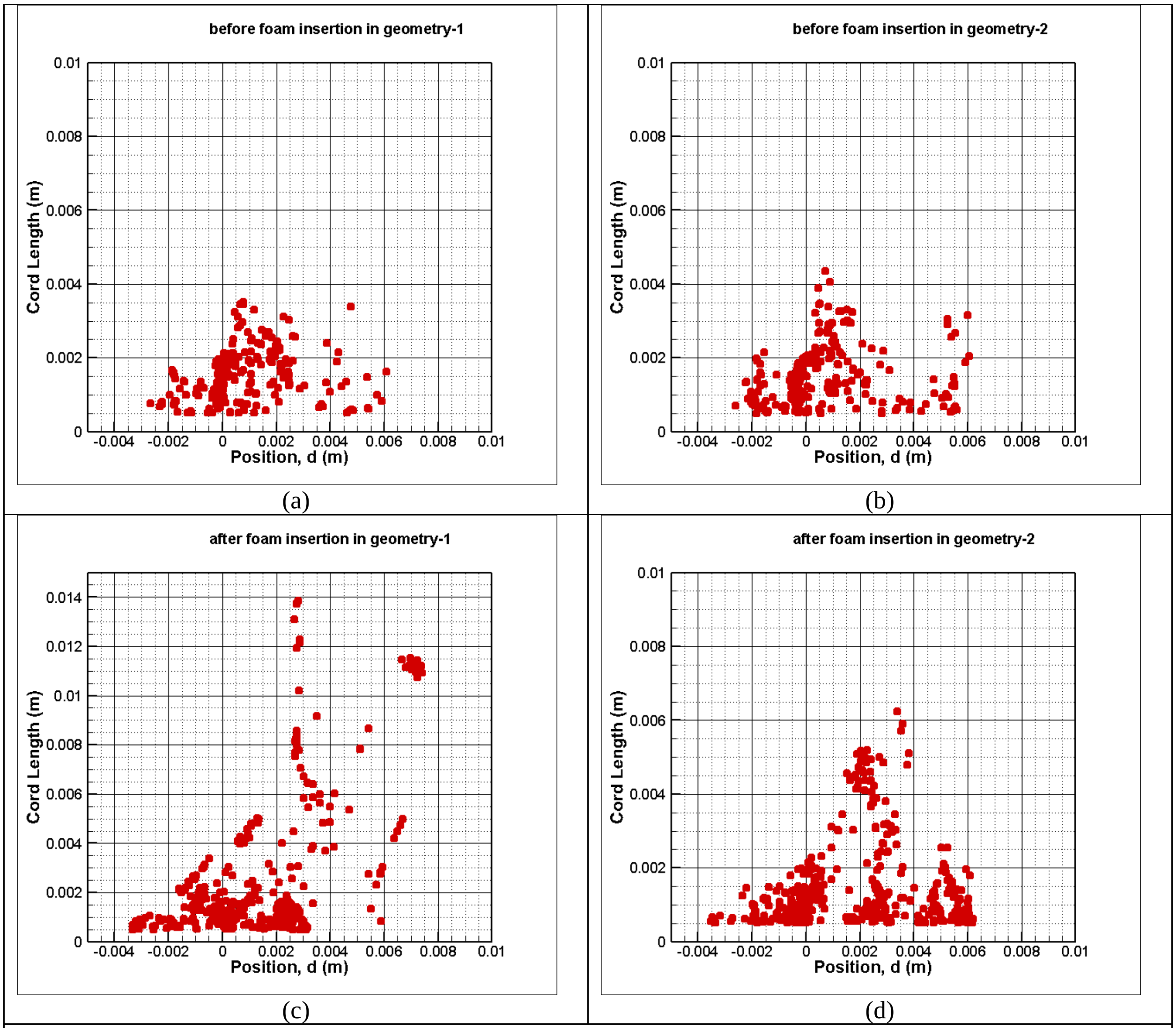

Figure 6: Vortex core lines' position and cord-length scatterplot for (a) before foam insertion in geometry-1 (b) before foam insertion in geometry-2 (c) after foam insertion in geometry-1 (d) after foam insertion in geometry-2

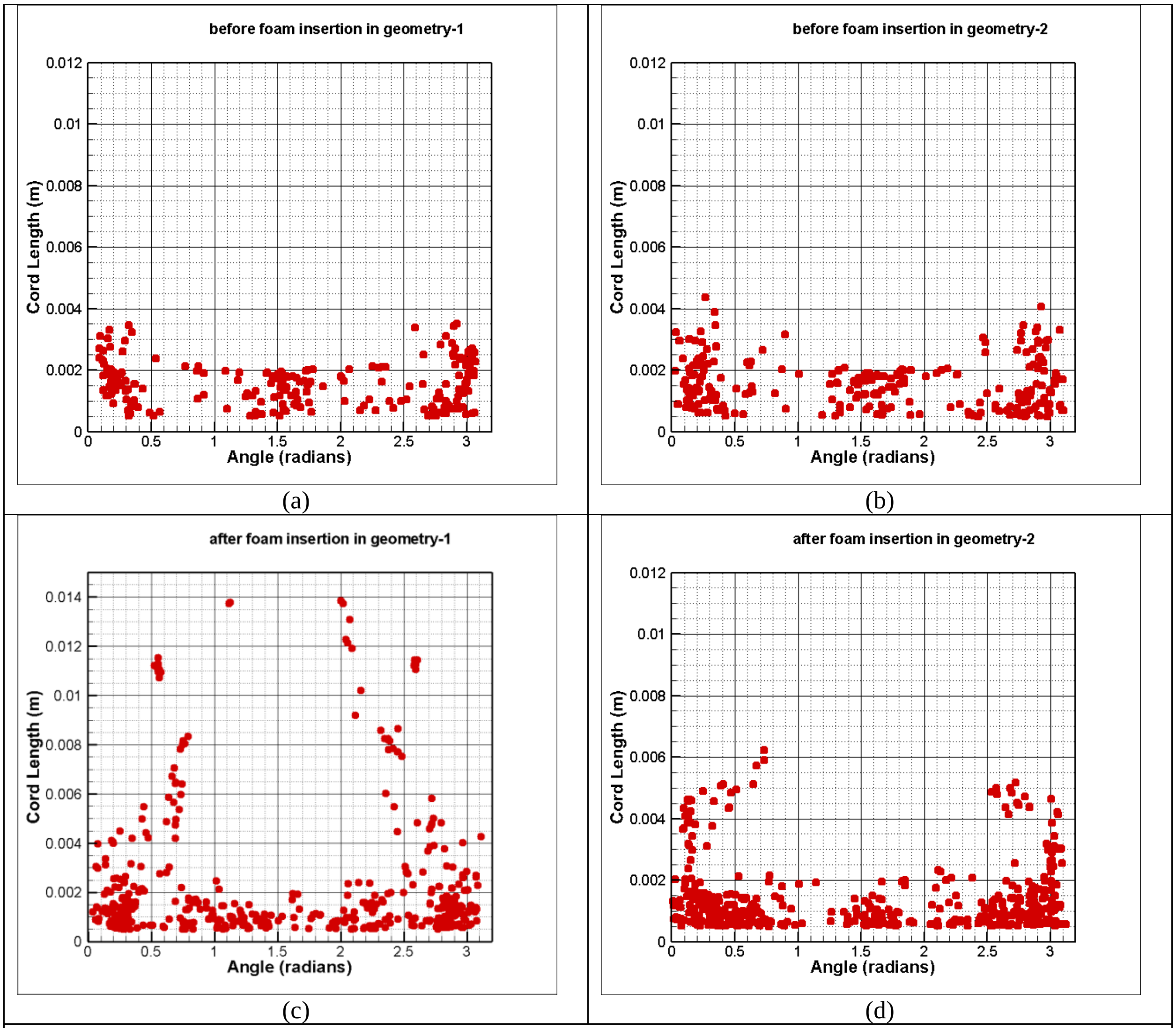

Figure 7: Vortex core lines' angle and cord-length scatterplot for (a) before foam insertion in geometry-1 (b) before foam insertion in geometry-2 (c) after foam insertion in geometry-1 (d) after foam insertion in geometry-2

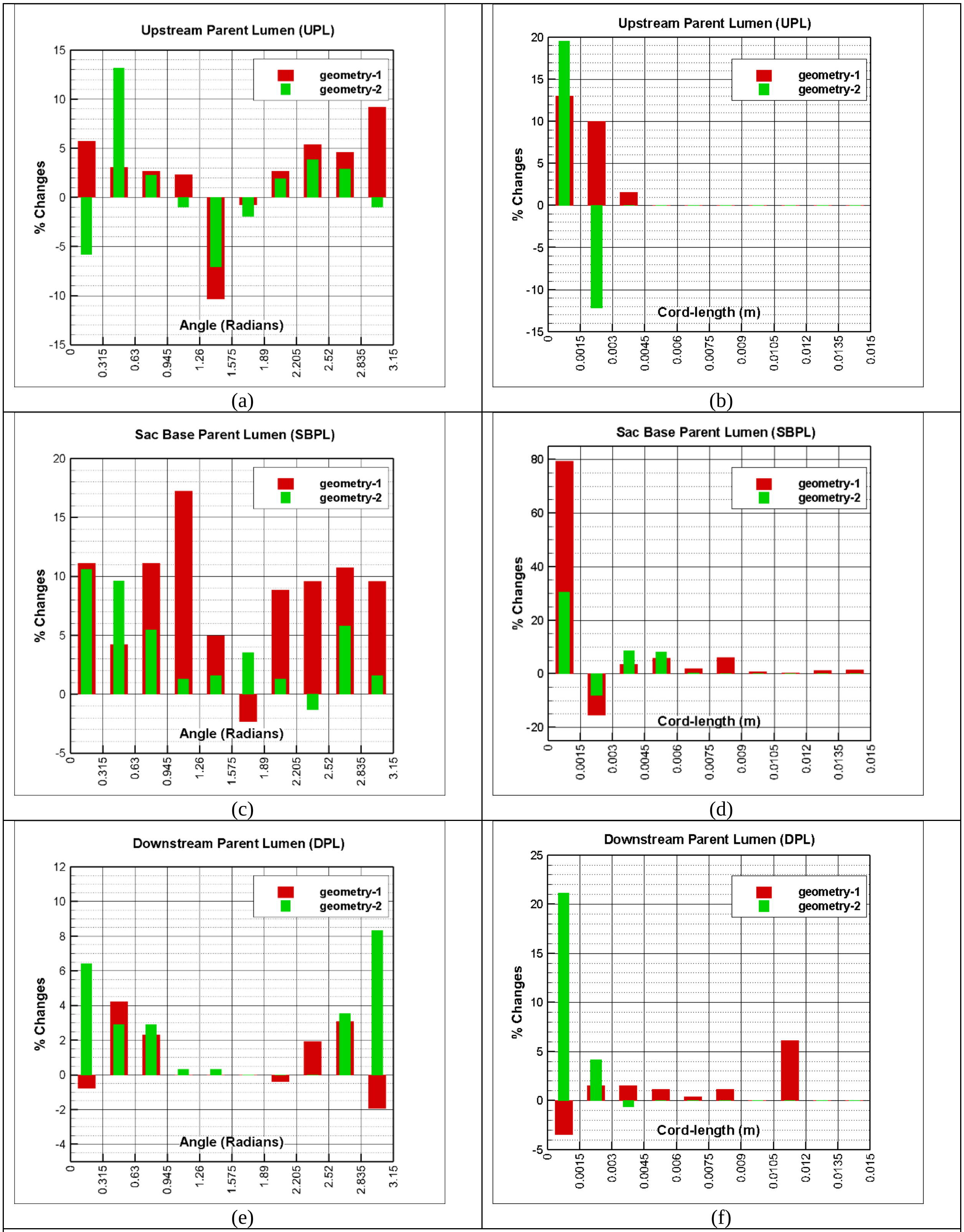

Figure 8: Histogram of change in vortex core line's (a) angle in upstream parent lumen (UPL) (b) cord-length in UPL (c) angle in sac base parent lumen (SBPL) (d) cord-length in SBPL (e) angle in downstream parent lumen (DPL) (f) cord-length in DPL. Changes are reported as the percentage of the total number of vortex core lines (VCLs) present in the corresponding segment before foam insertion

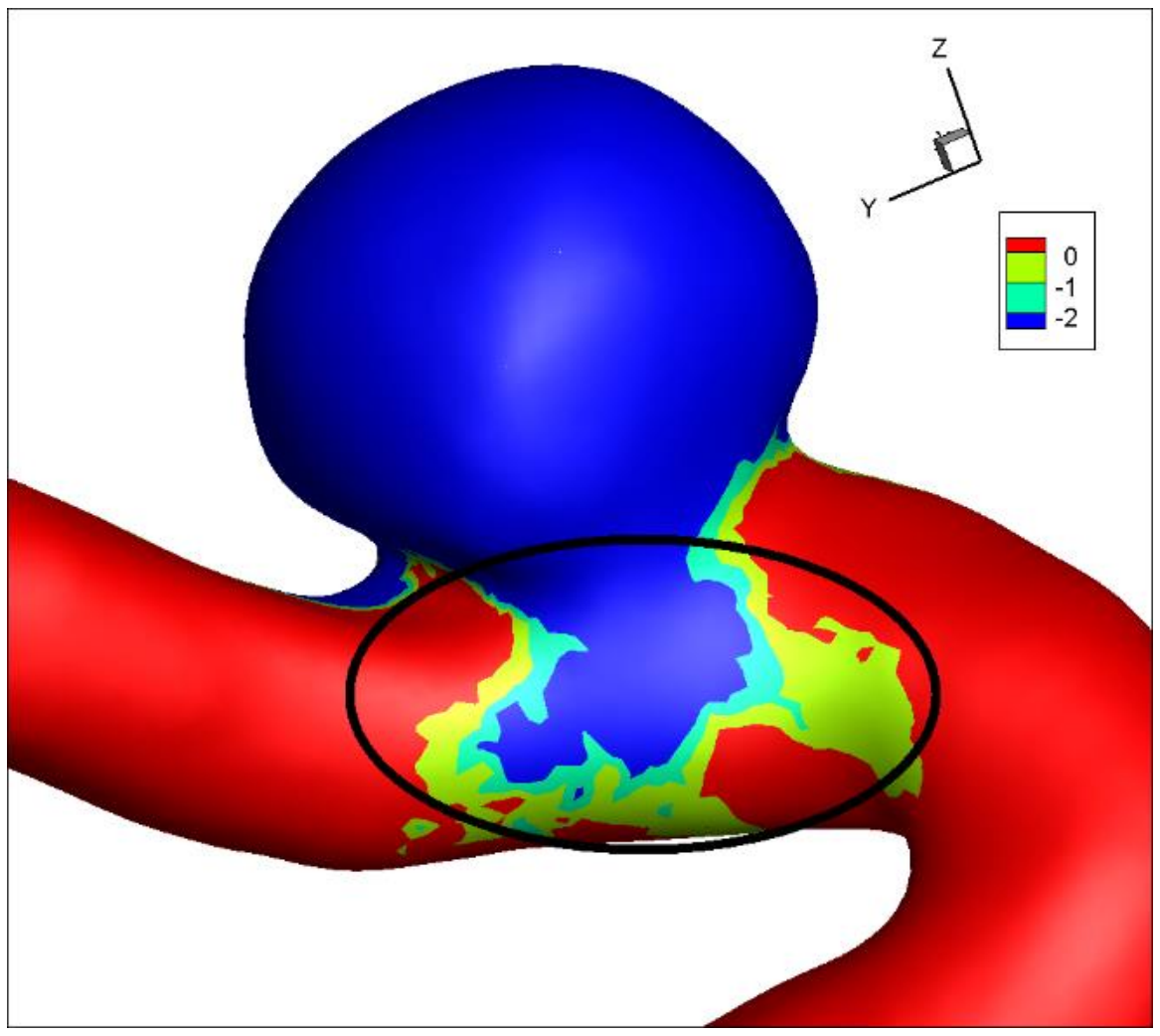

(a)

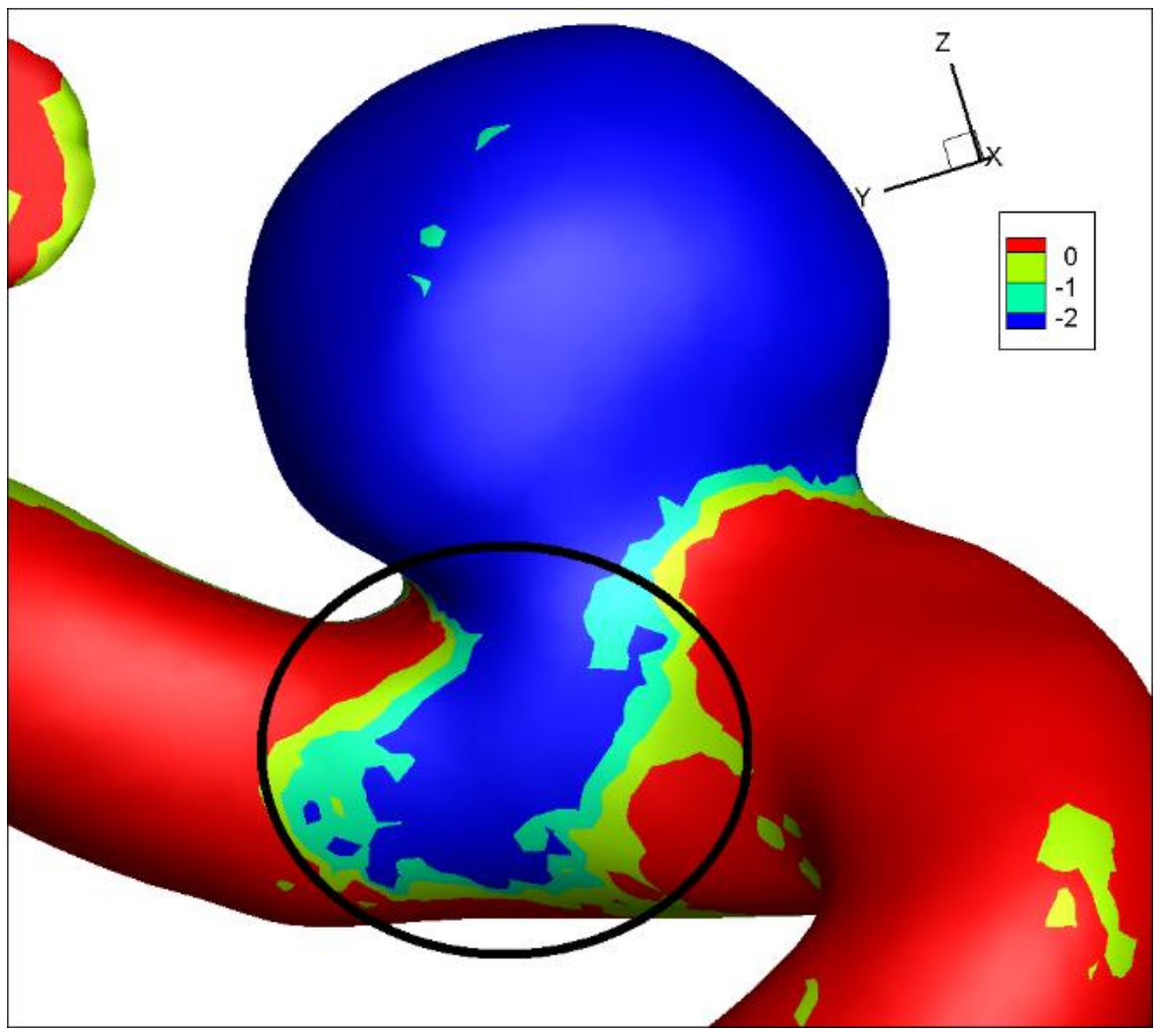

(b)

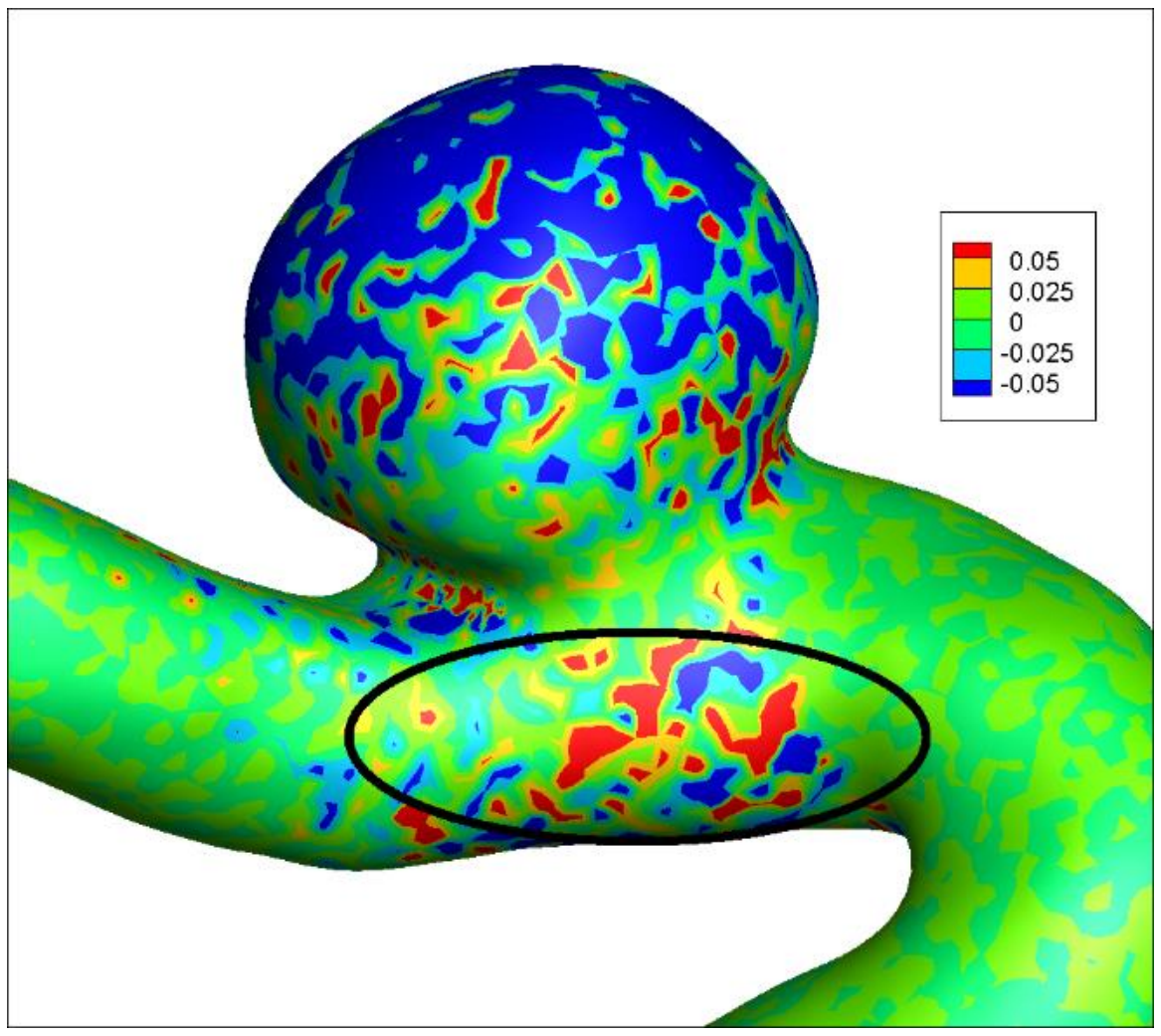

(c)

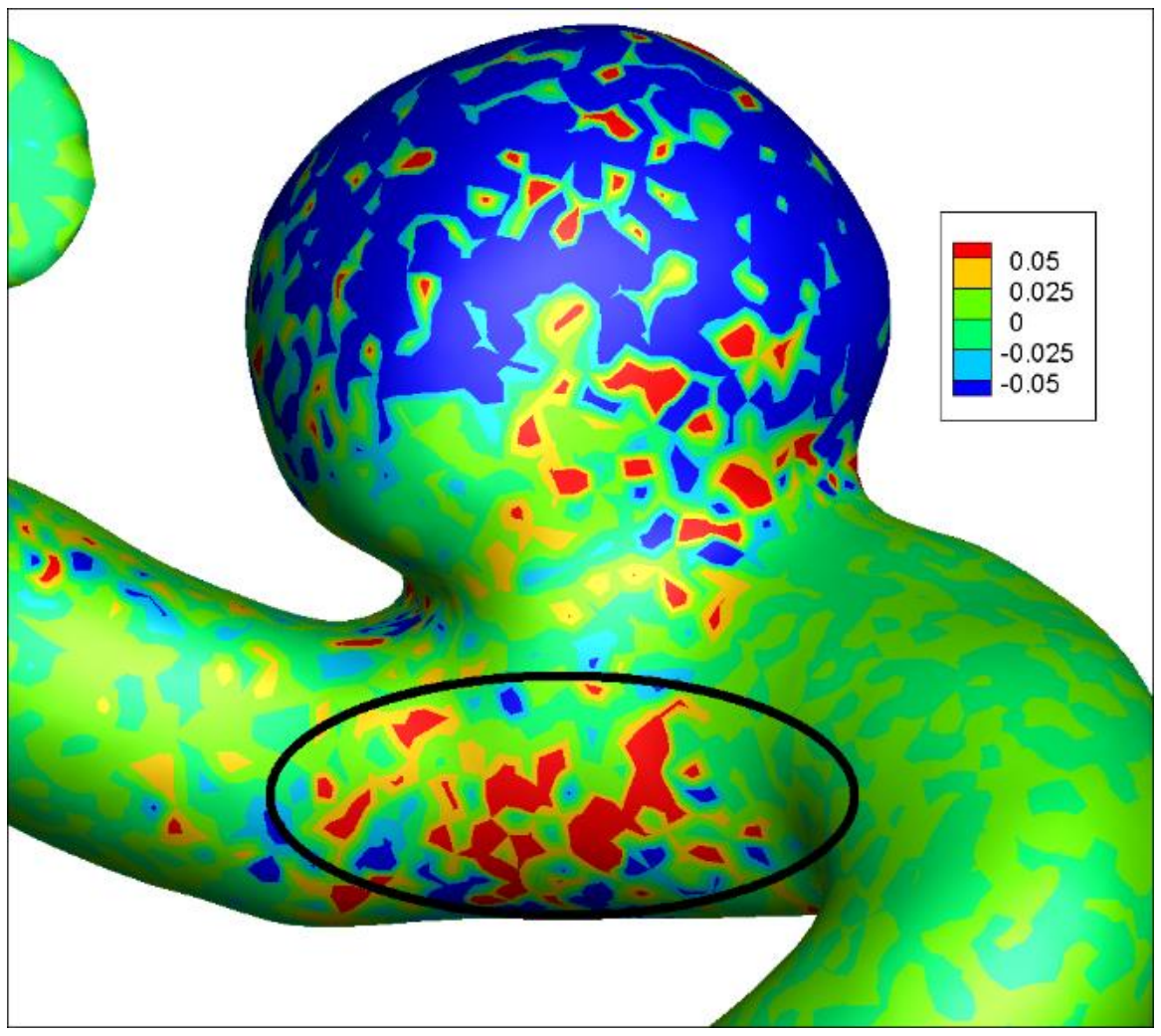

(d)

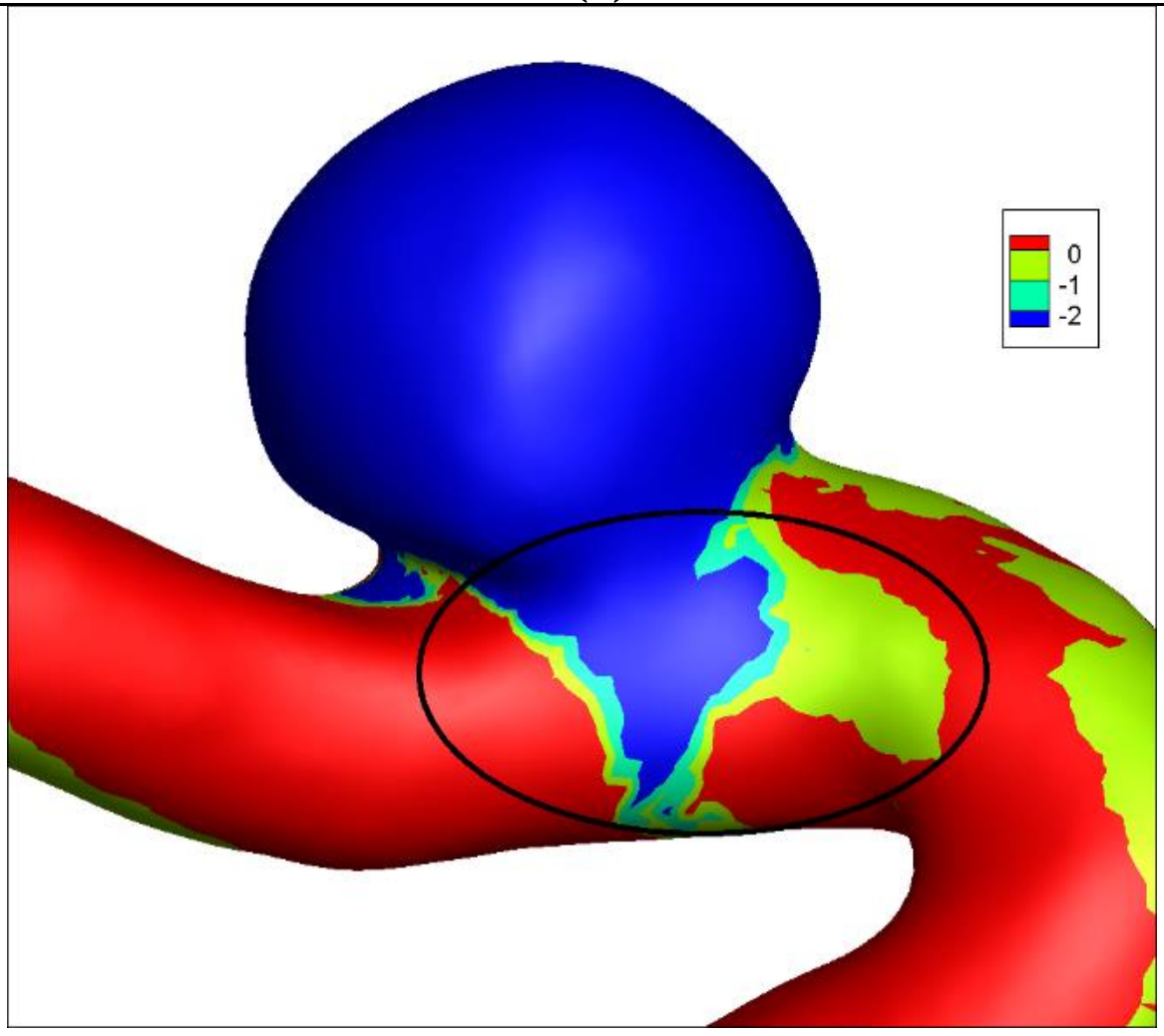

(e)

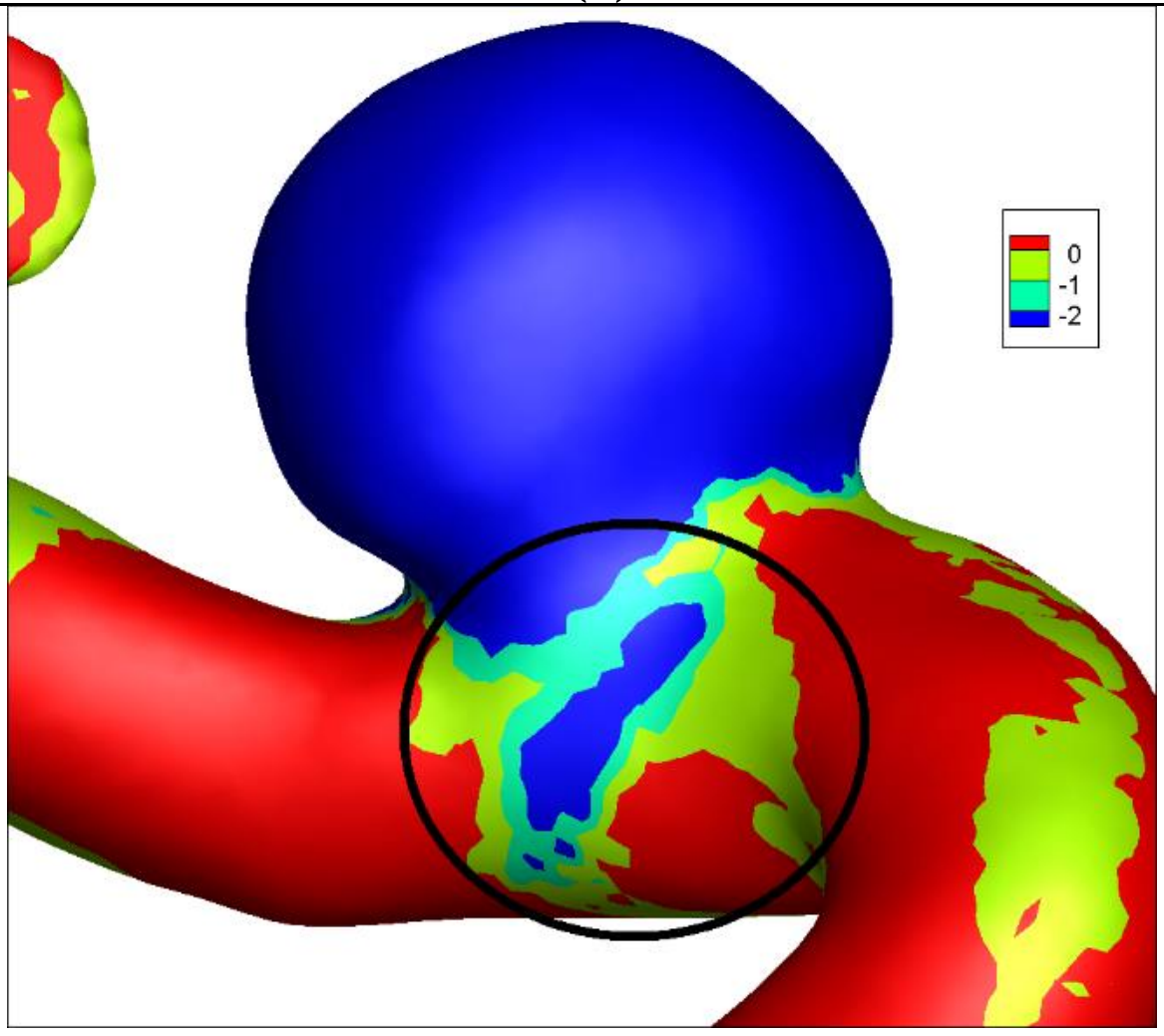

(f)

Figure 9: Contours of change in time-averaged wall shear stress (ΔTAWSS) in (a) geometry-1 (b) geometry-2; encircled region show reduction in time-averaged wall shear stress (TAWSS) due to foam insertion; Contours of change in oscillatory shear index (ΔOSI) in (c) geometry-1 and (d) geometry-2; The encircled region shows patches of increase in the oscillatory shear index (OSI) due to foam insertion; Contours of change in time-averaged secondary wall shear stress (ΔTASWSS) in (e) geometry-1 (f) geometry-2; Except for the aneurysm sac, the sac-base region (encircled region) is the only part of geometry to experience reduced time-averaged secondary wall shear stress (TASWSS)

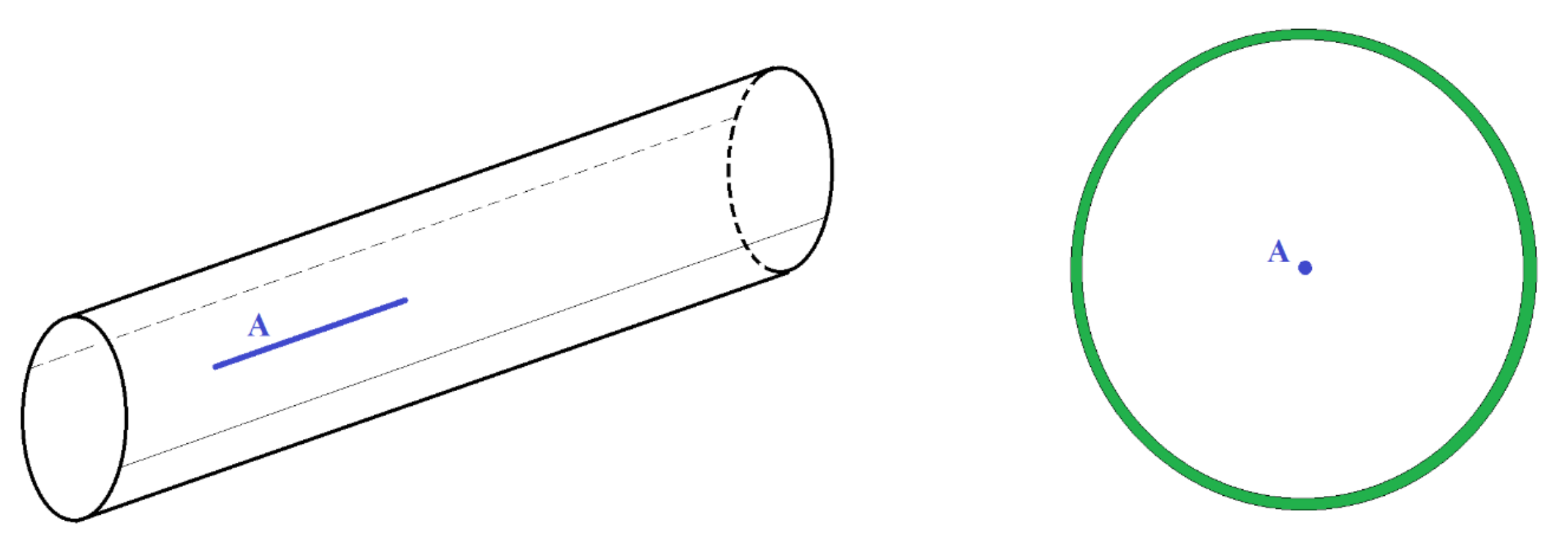

(a)

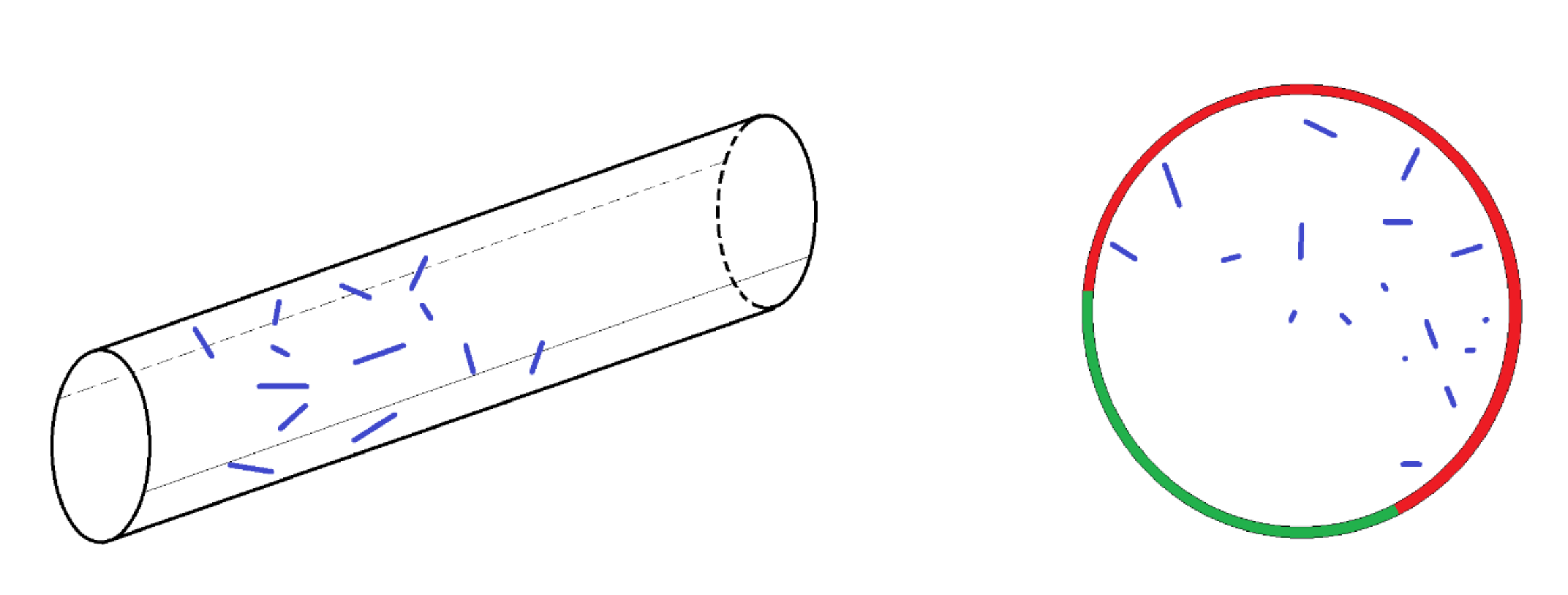

(b)

Figure 10: Schematic illustration of vortex core line (VCL) orientations and their exposure to the wall (a) single VCL in lumen causing physiologically positive effects (b) numerous oblique and small VCLs causing adverse exposure on the arterial wall. Isometric view (on the left) and cross-sectional view (on the right) are shown for both arrangements. The red region on the walls represents disturbed shear, i.e., adverse exposure on the wall. The green region represents the benign exposure of vortex structures. VCLs are shown in blue color